\documentstyle{article}
\begin{document}
\pagestyle{empty}
\parindent=1cm
\parskip=0.5cm
\vspace*{2.0cm}
\begin{center}
{\bf Parametric ordering of complex systems}\\
\vspace{1.0cm}
P.-M. Binder\\
\vspace{1cm}
IIASA, A-2361 Laxenburg, Austria\\
\vspace{0.5cm}
and\\
\vspace{0.5cm}
Wolfson College, Oxford OX2 6UD, United Kingdom\\
\end{center}
\vspace{1cm}
\noindent
Cellular automata (CA) dynamics are ordered in terms of two
global parameters, computable {\sl a priori} from the description
of rules. While one of them (activity) has been used before,
the second one is new; it estimates the average
sensitivity of rules to small configurational changes.
For two well-known families of rules, the Wolfram
complexity Classes cluster satisfactorily.
The observed simultaneous occurrence of
sharp and smooth transitions from ordered to
disordered dynamics in CA can be explained with
the two-parameter diagram.

\vspace{1cm}
\noindent
{\bf PACS numbers:} 05.45.+b, 02.60.-x, 05.70.-a, 89.80.+h

\vfil\eject
\pagestyle{plain}
\setcounter{page}{1}
In recent years cellular automata (CA) have played two important
roles. As totally discrete models (in space, time and
state variable) they have
proved very useful for the computer simulation of diverse spatially
extended problems in physics, chemistry and biology$^{1-6}$.
They have also been an important
tool in the classification of complex spatio-temporal
behavior$^7$ and in the understanding of the origins of complexity
in physical systems$^8$.
In this paper we address the related problem of {\sl
ordering} CA rules in terms of parameters that can be calculated
{\sl a priori}
from the description of a rule, without performing simulations.
We do this in terms of the well-studied activity parameter$^9$, described
below, and of a new parameter which estimates
average sensitivity to small changes in configuration. The two
main results are that rules cluster satisfactorily in the two-parameter
diagram according to their Wolfram Class$^{7(a)}$, and that
the diagram can explain
the simulational observation$^{9}$ that either first or second-order
transitions from ordered to disordered dynamics can be observed over
the same range of the activity parameter.

This work concentrates on
one-dimensional, two-state ($k=2$) {\sl totalistic} rules with
radius $r=2$. For these, sites which can take on the values
zero or one are updated simultaneously at integer time steps
based on deterministic functions $f$ of the sum of
the five neighbors. Upon defining zero as a quiescent state
(the outcome of the neighborhood $00000$ is $0$),
32 rules remain in this family.
Results for elementary CA ($k=2, \ r=1$)
will also be sketched here, and will be
presented in detail elsewhere$^{10}$.
These two families have been chosen because simulations
of their behavior
are quite well documented (Ref. 3).
The four Wolfram Classes are represented in this set of
rules. They are: (1) evolution leading to a homogeneous
state (all zeros or ones); (2) evolution leading to a set of separated
simple or periodic structures; (3) evolution leading to chaotic patterns,
and (4) evolution leading to complex localized structures, sometimes
long-lived. Examples of these classes can be found in Ref. 3.
Classes 1 and 2 are considered to be ordered, Class
3 is chaotic, and Class 4 is somewhere in between, exhibiting
complex structures that live for very long times$^{11}$.

Langton and coworkers $^9$ identified a global parameter that
turns out to be a useful predictor of
the complexity of a CA rule. Their activity parameter
$\lambda$
is the fraction of non-zero outputs of the CA
transition function $f$ when all possible arguments
(neighborhoods) of $f$ are weighted equally. The activity parameter
($0 \leq \lambda \leq 1$) estimates roughly the asymptotic
fraction of active (non-zero) sites.
Clearly, rules where most sites go to zero or to one
very quickly tend to homogeneous global states (Class 1),
while on the other hand at intermediate values of activity there
are more global states to be explored and one would expect
more disorder$^{12}$.
In Refs. 9, several effectively continuous
measures of complexity$^{13}$ were
monitored as $\lambda$ was increased.
A correlation between complexity and activity was
found, but it has become
clear that this parameter alone cannot predict the dynamics of a
rule. For example,
the identity rule, which freezes in time all initial conditions,
has $\lambda=1/2$ and yet is quite ordered.
When the number of states or neighborhood size is large enough
that $\lambda$ is suitably fine-grained,
the order-to-chaos transition with increasing activity
sometimes happens gradually (reminiscent of a second-order transition,
usually via Class 4 rules), and other times suddenly, as in a
first-order phase transition$^9$. This points to the need for additional
parameters. Attempts have been made by defining generalized versions of
thermodynamic quantities or quantities derivable from mean-field
theories$^9$, but no satisfactory answers have emerged yet.

To address this need
we have used a new sensitivity parameter ($\mu$), motivated by
the numerical observation that the Wolfram Classes
are not only characterized by their spatio-temporal patterns, but
also by their sensitivity to changes of the values of one or a few
sites (see Ref. 7(a) for example).
We therefore define the sensitivity parameter as the fraction
of changes in the evolution caused by changing the value of a site,
averaged over all sites of all neighborhoods in the rule table:
\begin{equation}
\mu = \frac{1}{nm} \sum\limits_n \sum\limits_{j=1}^m
\frac{\partial f}{\partial s_j},
\end{equation}
where $n$ are the possible neighborhoods in the rule table,
and $m$ is the number of bits in the neighborhood. The
Boolean derivative for CA$^{14}$ is
${\partial f}/{\partial s_j}=1$ if $f(s_1,\ldots ,s_j,\ldots)
\ne f(s_1,\ldots ,-s_j,\ldots)$,
this is, if the value of $f$ is sensitive to the value of
the bit $s_j$, and zero otherwise.
The minus sign is taken to mean bit complementation.
This estimates the average sensitivity to changes in a rule in
the same spirit as $\lambda$ estimates the asymptotic
concentration of active sites, and
is in some sense analogous to a spatial Lyapounov exponent.
A crucial point of this work is that average sensitivity can
be estimated {\sl a priori} and used as a {\sl predictor} of
complexity, instead of measured from simulations and used
as a {\sl measure} of complexity as has been done in Ref. 15
for the average difference pattern spreading rate.

A two-parameter diagram of totalistic ($k=2, \ r=2$) rules
is shown in Figure 1. It is seen that (1) all
Class 1 rules cluster at low ($\lambda, \mu$), (2) Class 2 rules
are contained in two separate clusters, at intermediate ($\lambda,
\ \mu$), (3) Class 4 rules occupy a high-sensitivity,
medium-activity
region and (4) Class 3 rules are quite prevalent and occupy much
of the parameter space. The clustering of rules by Wolfram Class
into distinct
portions of rule space is satisfactory, but several
comments are in order. First, the line between ordered rules
and chaotic rules is sometimes as sharp as the coarse-graining,
$\Delta \lambda=1/32$, allows; this indeed suggests
possible first-order dynamical phase transitions as activity
is increased. Second, the isolated bubble of Class 2 rules at
higher activity was expected$^9$, and corresponds in this case
to rules in which neighborhoods of sum 3 and 4 (and 5) produce
ones at the next time step.
The resulting frozen patterns of thick bands of ones
and zeros are typical of Class 2. Third, we have folded
the diagram along the line $\lambda=1/2$, as low and high activity
are dynamically equivalent; the rules for $\lambda > 1/2$
are shown in empty rather than full symbols.
Fourth, the arrows A and B demonstrate how, for the same value
of activity, one can observe either sharper or smoother
transitions depending on the sensitivity parameters
of the rules. Fifth, as the sensitivity parameter
increases with $\lambda=1/2$ fixed,
we see the expected sequence of Classes (1-2, 4 and 3).
Sixth, the Class 3 rule with very high activity and low
sensitivity which folds near the bottom left corner of the diagram
is atypical$^{16}$.
Seventh, the subclass of totalistic, quiescent rules may not
be indicative of the larger space of $2^{32}$ ($k=2, \ r=2$)
rules because of special symmetries.
Finally, the activity-sensitivity diagram for
the 88 independent {\sl elementary} CA looks quite similar,
but it only has three convex domains for Classes 1, 2 and 3.

The results in this paper help us understand the space of CA rules,
which can exhibit
spatio-temporal patterns ranging from ordered to complex to chaotic.
In particular, we have proposed a new parameter, which
estimates {\sl a priori} the average
sensitivity of rules to small perturbations.
The resulting diagrams in parameter space show that the rules
cluster according to their Wolfram Classes.
In fact, $\mu$ alone appears to be a better predictor of
complexity than $\lambda$.
There are exceptions, such as the Class 2 pocket with
higher $\lambda, \mu$, which remind us that such simple
parameters do not capture all the subtleties of rule
dynamics. Exceptions probably exist even with larger $k, \ r ,$
and dimension. The diagrams also illustrate
how different types of dynamical phase transitions can occur over
the same range of the activity parameter.
It is encouraging that the two parameters work well
in low-dimensional, small-radius rules, where correlations are
known to be the worst; we are currently investigating
the rule space of higher-dimensional CA and
hope that this work will be extended to study
the relation between the parameters presented here and those used to
describe continuous-variable complex systems such as coupled-map
lattices.

Two final remarks are in order. One is that $\mu$ may be refined
by separating the sensitivity to the central and peripheral sites
(corresponding roughly to information generation and transmission);
this is currently being investigated. The second remark is that
ordering in terms of such simple parameters is heuristic rather
than rigorous: the use of the activity parameter has recently been
criticized$^{17}$. However, careful examination of the rules that
escape simple predictions can lead to increasingly sharp definitions
of difficult concepts such as complexity.

The author thanks the French Association for the
Development of Systems Analysis for financial support,
and R. Bagley, C.G. Langton, W. Li, M. Mitchell,
K. Sigmund, and A. Wuensche for suggestions.

\vfil\eject

\vfil\eject

{\bf Figure 1.} Phase diagram of rules for totalistic $(k=2, \ r=2)$
cellular automata. Horizontal axis: activity parameter. Vertical
axis: sensitivity parameter. Circles: ordered rules (Wolfram
Classes 1 and 2, with Class 2 rules encircled and labeled C2);
Squares: complex rules (Wolfram class 4); Triangles: disordered
or chaotic rules (Wolfram class 3). Open symbols correspond to rules
with large $\lambda$, which have been reflected about the $\lambda =
1/2$ axis. The arrows correspond to a first-order phase transition (A),
and to a second-order phase transition (B). Any symbol may correspond
to more than one rule.


\begin{thebibliography}{990}
\bibitem{13}
For example, site or block entropies and mutual information
are particularly useful measures. Both are used in
Ref. 9.
\end{thebibliography}
\end{document}